\documentclass[12pt]{article}

\usepackage{amsmath,amssymb,amsthm,epsfig,amscd}

\numberwithin{equation}{section}       
\newcommand{\cH}{{\cal{H}}}

\newcommand{\ket}{\rangle}
\newcommand{\bra}{\langle}
\newtheorem{theorem}{Theorem}
\newtheorem{lemma}{Lemma}

\begin{document}


\begin{titlepage}

\vfill

\begin{flushright}
\begin{small}
   Imperial/TP/2006/DR/01\\
   \end{small}
\end{flushright}

\vfill
\vfill

\begin{center}
   \baselineskip=16pt
   \begin{LARGE}
      \textsl{Popescu-Rohrlich Boxes\\*[0.6em]
        in Quantum Measure Theory}
   \end{LARGE}
   \vskip 2cm
      Matthew Barnett, Fay Dowker and David Rideout
   \vskip .6cm
   \begin{small}
      \textit{Blackett Laboratory, Imperial College\\
        London, SW7 2AZ, U.K.}
        \end{small}
\end{center}

\vfill

\begin{center}

Dedicated to Rafael Sorkin on the occasion of his 60th birthday. 
\end{center}
\vfill

\begin{small}
\begin{center}
   \textbf{Abstract}
\end{center}
Two results are proved at the quantal level in Sorkin's
hierarchy of measure theories. One is a strengthening
of an existing  bound on the correlations in the EPR-Bohm setup
under the assumption that the probabilities admit a
strongly positive joint quantal measure.
 It is also proved that any set of
no-signalling probabilities, for two distant
experimenters with a choice of two
alternative experiments each
and two possible outcomes per experiment,
admits a joint quantal measure, though one that is not
necessarily strongly positive.

\begin{quote}
\vfill
\vfill

\end{quote}
\end{small}

\end{titlepage}


\section{\label{sec:intro}Introduction}

An answer to the question ``What is the essential nature 
of a quantum theory?'' has been proposed by Rafael Sorkin. 
A quantum theory, according to Sorkin's scheme of 
generalised measure theories, is one in which 
pairs of alternative spacetime histories of a system 
can interfere but in which there is 
no interference among triples of histories that is not 
already accounted for by 
the pairwise interference \cite{Sorkin:1994dt}. 
All quantum theories, by this criterion, are placed 
at the second level in a hierarchy of measure theories: a level 
$k$ theory allows interference among $k$-tuples of 
histories but no irreducible interference among $(k+1)$-tuples.
Classical stochastic theories (and deterministic
theories as special cases) are contained at level 1. 

The surprising thing about this scheme is that 
the class of quantum theories
includes all standard unitary quantum theories but is 
by no means exhausted by them. There are
generalisations of the standard theory which 
yet remain at level 2.  
Even more surprisingly, one of these generalisations has 
been shown \cite{Craig:2006} to 
include the device that has come 
to be known as the ``Popescu Rohrlich (PR) box'' \cite{Khalfin:1985, 
Popescu:1994}. 

Our purpose here is to investigate 
the structure of quantum measure theory including the 
way that PR boxes fit into it. We will concentrate on  
the situation we refer to as
the (2,2,2) case  which is   
the standard EPR-Bohm experimental setup in which the 
Clauser-Horne-Shimony-Holt-Bell (CHSHB) inequalities are derived: 
2 distant experimenters Alya and Bai, each have a choice of 
2 alternative 
settings for their experiment and each experiment yields one of 
2 possible macroscopic 
outcomes. 

In section 2 we review the basic aspects of Sorkin's hierarchy of measure theories. We focus on level 2,
the quantal level, which is closely related to Hartle's generalised
quantum mechanics \cite{Hartle:1989,Hartle:1992as} without the
requirement of decoherence or ``consistency''. 
We review
the properties of a decoherence functional
and define the concept of strong positivity for
quantal measures/decoherence functionals.
In section 3 we show that there is
a set of inequalities on the correlations
that is implied  by the existence of a strongly positive
joint decoherence functional or quantal measure on the (fictitious)
joint sample space of all possible outcomes of both of Alya's and
both of Bai's experiments.
This tightens the bounds derived in
reference \cite{Craig:2006}.

In section 4 we prove our main result that {\it{any}} system of
no-signalling 
experimental probabilities in our (2,2,2) set-up admits a 
joint quantal measure. 
In section 5 we conclude with tentative conjectures on more complicated
set-ups, the corresponding PR-type boxes and 
higher levels in the Sorkin hierarchy.  

\section{Generalised measure theory} 

We introduce the hierarchy of measure theories and refer to
references \cite{Sorkin:1994dt,Sorkin:1995nj,Salgado:1999pu} 
for details.
In a generalized measure theory, there is a sample space $\Omega$ of
possible histories for the system in question.
Normally these are  to be thought of as ``fine grained histories'',
meaning as complete a description of physical reality as is conceivable
in the theory, {\it e.g.} for $n$-particle mechanics a history would be
a set of $n$ trajectories, and for a scalar field theory, a history
would be a field configuration on spacetime.
Predictions about the system --- the dynamical
content of the theory ---  are to be gleaned,
in some way or another, from
a (generalized) measure $\mu$, a non-negative real function on 
subsets of $\Omega$ 
(strictly, on some suitable class of ``measurable'' subsets of
$\Omega$). 

Given the measure,  we can construct the
following series of symmetric set functions:
\begin{align*}
I_1(X)  \equiv&\; \mu(X) \\
I_2(X,Y)  \equiv&\; \mu(X\sqcup Y) - \mu(X) - \mu(Y) \\
I_3(X, Y, Z)  \equiv&\; \mu(X\sqcup Y \sqcup Z) -
\mu(X\sqcup Y) - \mu(Y\sqcup Z) - \mu(Z\sqcup X) \\
	      &+ \mu(X) + \mu(Y) + \mu(Z) 
\end{align*}
and so on, where $X$, $Y$, $Z$, {\it etc.\ }are disjoint subsets of $\Omega$,
as indicated by the symbol `$\sqcup$' for disjoint union.

A {\it measure theory of level $k$} is defined as
one which satisfies the sum rule 
$I_{k+1}=0$.  
This condition implies that all higher sum rules are 
automatically satisfied, {\it viz.\ }$I_{k+n}=0$ for all $n\geq 1$.
This means that each level of the hierarchy contains all the 
levels below it. 
A level 1 theory is 
one in which the measure satisfies the usual
Kolmogorov sum rules of classical probability theory, 
classical
Brownian motion
being a good example.  
A level 2 theory is one in which the Kolmogorov sum rules
may be violated but $I_3$ is nevertheless zero.  Unitary quantum mechanics
satisfies this condition and so we call level 2 generally
the level of quantum measure theory.  

The existence of a 
normalized 
quantal  measure on $\Omega$ is equivalent \cite{Sorkin:1994dt} to the
existence of a 
{\it decoherence functional} 
$D(X;Y)$ of pairs of subsets of $\Omega$
satisfying \cite{Hartle:1989,Hartle:1992as}:

\noindent (i) Hermiticity: $D(X;Y) = D(Y;X)^*$ ,\  $\forall X, Y$;

\noindent (ii) Additivity: $D(X\sqcup Y; Z) = D(X;Z) + D(Y;Z)$ ,\
$\forall X, Y, Z$ with $X$ and $Y$ disjoint;

\noindent (iii) Positivity: $D(X;X)\ge0$ ,\  $\forall X$;

\noindent (iv) Normalization: $D(\Omega ;\Omega)=1$  .

\noindent The quantal measure is given by  
a decoherence functional via 
\begin{equation}
\mu(X) = D(X ; X).       
\end{equation}
{The quantity $D(X;Y)$ is 
interpretable as the quantum interference between two sets of
histories in the case when they are disjoint.  
Notice 
that $\mu$
determines only the real part of $D$.} To see how the 
decoherence functional is defined for ordinary non-relativistic
particle quantum mechanics see for example \cite{Hartle:1992as,
Hartle:2006nx}. 

A further condition on a decoherence functional 
is the condition of {\it strong positivity}, 
which states that for any finite collection of
(not necessarily disjoint) 
subsets $X_1,X_2,\dots X_n$
of $\Omega$,
the $n \times n$
matrix $M_{ij} \equiv D(X_i\,; X_j)$ is positive
semidefinite. 
The decoherence functional of any ordinary unitary quantum mechanical
theory is strongly positive. So is the decoherence functional 
for any open quantum system that is derived from a unitary 
model by summing out over some ignored variables. 
The only concrete dynamical decoherence functionals 
known to us that are not derivative of an underlying unitary 
model are those for the family of ``quantum random walks'' 
of Martin et al \cite{Martin:2004xi} and these are also strongly 
positive.  

Strong positivity 
is a powerful requirement because it
implies 
that there is a Hilbert space associated with the quantum
measure, which turns out to be the standard Hilbert space in the case of
unitary quantum mechanics \cite{Martin:2004xi,Dowker:2004}.
Strong positivity plays an important role in the present 
investigation as will be made clear. Indeed it will be 
shown that the question of why there are (apparently) no PR boxes
in nature becomes, in the quantum measure theory framework, 
the question of why the decoherence functional is strongly 
positive. 

We will not enter into
the general question of how to
interpret the quantal measure. 
Suffice it to mention that  
one set of ideas for doing so
goes by the name of ``consistent histories'' or ``decoherent histories''
and attempts in effect to reduce the quantal measure to a classical one
by the imposition of decoherence conditions \cite{Griffiths:1984rx,
Omnes:1988ek, Gell-Mann:1989nu,Hartle:1992as}.
And a different attempt at an interpretation, 
in which the microscopic world is just as real as the 
macroscopic, may be found in \cite{Sorkin:1995nj}.
For our purposes in this paper,
it will be enough to assume,
where macroscopic measuring instruments are concerned,
that distinct ``pointer readings'' do not interfere 
and 
that their quantal measures can 
be interpreted as probabilities in the sense of frequencies.

\section{The Tsirelson inequalities, I, II and III}

We follow \cite{Craig:2006} in setting up the following notation.
Consider two distant experimenters, Alya and Bai, who each perform 
one of two possible experiments: Alya chooses setting $a$ or $a'$
for her experiment, and Bai chooses setting $b$ or $b'$ for his. 
Each then obtains one of two possible outcomes, $+1$ or $-1$. 
This means, that, a priori, 
one has 
four entirely distinct probability distributions, $\mathbb{P}_{\alpha\beta}$
each defined on its own four-element sample space
$\Omega_{\alpha\beta}=\Omega_\alpha\times\Omega_\beta$,
where 
$\alpha$ ranges over the settings $a$ or $a'$,
$\beta$ ranges over the settings $b$ or $b'$,
and 
each space $\Omega_\alpha$, $\Omega_\beta$ is a binary sample space,
corresponding to the two 
possible experimental outcomes, $\pm 1$. 
We refer to the four probability distributions, 
$\mathbb{P}_{ab}$, $\mathbb{P}_{ab'}$,
$\mathbb{P}_{a'b}$ and $\mathbb{P}_{a'b'}$,
together as a system of experimental probabilities. 
We will make the assumption throughout that 
 every system of experimental 
probabilities satisfies the ``no-signalling'' conditions, namely,
the marginal probabilities for the outcomes of Alya's $a$ experiment 
are the same whether calculated from 
$\mathbb{P}_{ab}$ or $\mathbb{P}_{ab'}$ and
similarly for her $a'$ experiment and similarly for Bai's experiments. 

Now we merge the $\Omega_{\alpha\beta}$ into 
a single sample space
\begin{equation*}
\Omega \equiv 
\Omega_a \times \Omega_{a'} \times \Omega_b \times \Omega_{b'}
\end{equation*}
of $2^4=16$ elements.
Let us label the elements of $\Omega$ by 
the sixteen 4-element bit strings $\{ (i i' j j'): i,i',j,j' = \pm 1\}$ 
where $i$ corresponds to the $a$ outcome, $i'$ to $a'$, $j$ to $b$ 
and $j'$ to $b'$.

We say that the experimental probabilities {\it admit 
a joint quantal measure} iff there exists a decoherence 
functional ${D}$ on $\Omega$ such that its marginals
agree with the experimental probabilities in the following 
way
\begin{align} 
\sum_{i'j'k'l'} {D}(i i' j j' \, ;\, k k' l l')
& =
\mathbb{P}_{ab}(ij) \delta_{ik} \delta_{jl}\ , \notag\\
\sum_{i'jk'l} {D}(i i' j j' \, ;\, k k' l l')
& = 
\mathbb{P}_{ab'}(ij') \delta_{ik} \delta_{j'l'}\ , \notag\\
\sum_{ij'kl'} {D}(i i' j j' \, ;\, k k' l l')
& = 
\mathbb{P}_{a'b}(i'j) \delta_{i'k'} \delta_{jl}\ , \notag\\
\sum_{ijkl} {D}(i i' j j' \, ;\, k k' l l')
& = 
\mathbb{P}_{a'b'}(i'j') \delta_{i'k'} \delta_{j'l'}\ .
\label{quantummarginals}
\end{align}

If there  exists such a joint decoherence functional that is 
moreover strongly positive we say that the experimental 
probabilities admit a strongly positive joint quantal measure
(SPJQM).  

Note that the marginal decoherence functionals, one for 
each realizable pair of experiments, are required to 
decohere on the outcomes.

It was shown in \cite{Craig:2006} that certain inequalities on the
correlations, known as the Tsirelson inequalities \cite{Tsirelson:1980},
follow from the existence of a SPJQM.
If we define
$X_{\alpha \beta}$ to be 
the ``correlator'' for the $(\alpha, \beta)$ experiment via:
\begin{align*}
X_{ab} = &\sum_{i j}\; i\; j\; \mathbb{P}_{ab} (ij)\\
X_{ab'} = &\sum_{i j'}\; i\; j'\; \mathbb{P}_{ab'} (ij')\\
X_{a'b} = &\sum_{i' j}\; i'\; j\; \mathbb{P}_{a'b} (i'j)\\
X_{a'b'} = &\sum_{i' j'}\; i'\; j'\; \mathbb{P}_{a'b'} (i'j')\;
\end{align*}
then the 
inequalities are 
\begin{equation*}\label{tsirelsoni}
|\; X_{ab} + X_{a'b} + X_{ab'} - X_{a'b'} \;| \le 2 \sqrt 2
\end{equation*} 
plus the three other inequalities where the minus sign is given to 
each of the other $X$'s in turn.  We will refer to these inequalities as TsirelsonI.
The CHSHB inequalities
have the same form except the $2\sqrt{2}$ is replaced by $2$. 

We  now improve these bounds 
to stronger inequalities.
First, as described in \cite{Craig:2006}, we 
apply to ${D}$ a certain basic construction
via which any strongly
positive decoherence functional
gives rise to a Hilbert space
\cite{Martin:2004xi,Dowker:2004}.
Indeed, if the experimental probabilities 
admit a SPJQM, via a decoherence functional 
${D}$ as above, then there is a Hilbert 
space $\cH$ spanned by vectors 
$\{|i i' j j'\rangle\}$ labelled by the elements of 
${\Omega}$ and on which 
the inner product is given by
\begin{equation} \label{D-braket}
\langle i \, i' \, j \, j' \vert k \, k' \, l \, l'\rangle  
={D}(i \, i' \, j \, j' ; k \, k' \, l \, l') \ .
\end{equation}

We will require the following Lemma, the proof of
which is given in \cite{Craig:2006}.

\begin{lemma}\label{lemma1}
Let the experimental probabilities admit a SPJQM
given by the decoherence functional ${D}$ on $\Omega$.
Let $|a\ket$, $|a'\ket$, $|b\ket$, and $|b'\ket$ $\in\cH$ be defined by
\begin{align} 
|a\ket &= \sum_{i i' j j'} i \,|i i' j j'\ket\, ,\notag \\
|a'\ket &= \sum_{i i' j j'} i'  |i i' j j'\ket\, , \notag\\
|b\ket &= \sum_{i i' j j'} j \, |i i' j j'\ket\, , \notag\\
|b'\ket &= \sum_{i i' j j'} j' |i i' j j'\ket\, . 
\label{a-ket}
\end{align}
Then
\begin{equation*}
\bra{a}|{a}\ket = \bra{b}|{b}\ket = 
\bra{a'}|{a'}\ket = \bra{b'}|{b'}\ket = 1
\end{equation*}
and
\begin{equation}\label{Xbrak}
\bra{a}|{b}\ket = X_{ab}\; , 
\bra{a}|{b'}\ket =  X_{ab'}\; , 
\bra{a'}|{b}\ket = X_{a'b}\; ,
\bra{a'}|{b'}\ket = X_{a'b'}\; . 
\end{equation}
\end{lemma}

We are now ready to prove our first result.

\begin{theorem}
If a 
system of experimental probabilities for the (2,2,2) system 
admits a SPJQM then 
\begin{equation}\label{tsirelsonii}
| \;\arcsin X_{ab} + \arcsin X_{a'b} + \arcsin X_{ab'} 
- \arcsin X_{a'b'} \;|  \le \pi  
\end{equation}
where each angle, $\arcsin X_{\alpha\beta}$, lies between 
$-\pi/2$ and $\pi/2$. 
\end{theorem}

\begin{proof}
It suffices to prove  
\eqref{tsirelsonii} without the absolute value signs
as one sees by
reversing the signs of  Bai's outcomes.

Defining the angles 
\begin{align*}
\theta_1 &= \frac{\pi}{2} - \arcsin{X_{ab}},\\ 
	  \theta_2 &= \frac{\pi}{2} -\arcsin{X_{ab'}},\\
	  \theta_3 &= \frac{\pi}{2} -\arcsin{X_{a'b}},\\
	  \theta_4 &= \frac{\pi}{2} -\arcsin{X_{a'b'}}
\end{align*}
one sees that by Lemma 1
\begin{equation}
\cos\theta_1 = <a|b>\; , \cos\theta_2 = <a|b'>\; , 
\cos\theta_3 = <a'|b> \; , \cos\theta_4 = <a'|b'>\; 
\end{equation}
and the inequality to be proved boils down to an 
inequality on angles between 4 unit vectors in a 
real vector space. 

\begin{figure}[h]
\begin{center}
\includegraphics[width=2.5in]{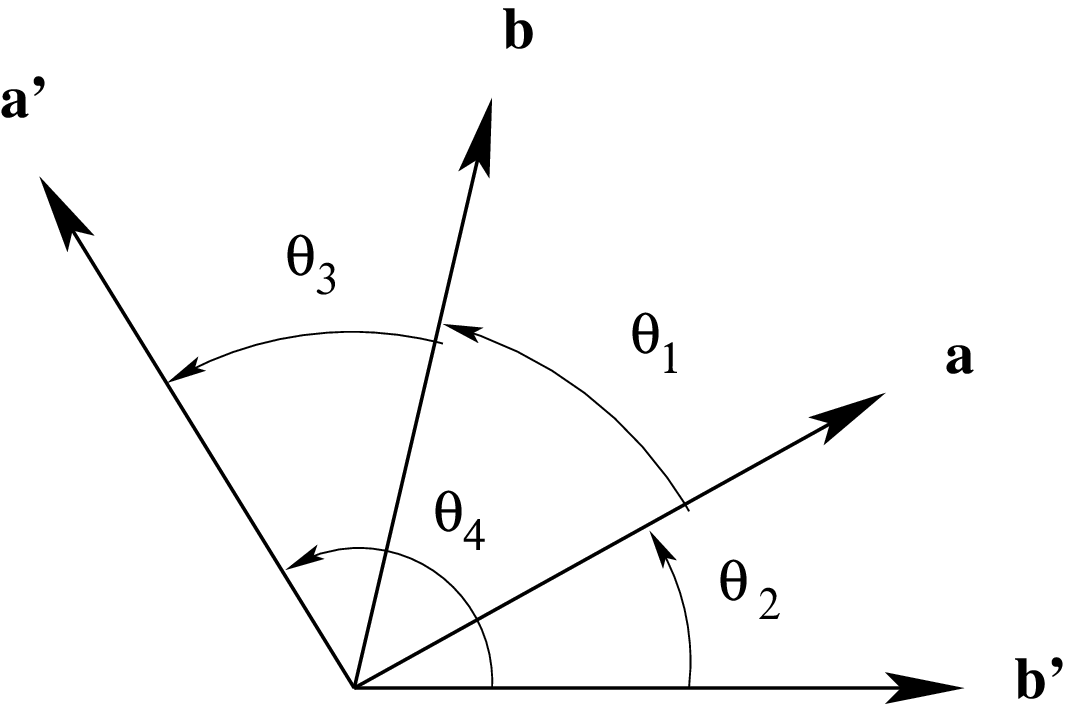}
\caption{Four coplanar vectors, ${\bf{a}}$, ${\bf{a'}}$, ${\bf{b}}$ and
${\bf{b'}}$.
}\label{fig:domain}
\end{center}
\end{figure}

Suppose $\theta_1 + \theta_2 + \theta_3 \le \pi$. Then 
if all four vectors are coplanar, $\theta_4 
= \theta_1 + \theta_2 + \theta_3$ (see figure 1) and if they are not 
coplanar, $\theta_4 \le \theta_1 + \theta_2 + \theta_3$.
If $\theta_1 + \theta_2 + \theta_3 > \pi$ then 
$\theta_1 + \theta_2 + \theta_3 - \theta_4 \ge 0$,
since $\theta_4 \le \pi$. 
In either case we have 
\begin{equation*}
-\theta_1 - \theta_2 - \theta_3 + \theta_4 \le 0
\end{equation*}
and hence the result.
\end{proof}
The other three inequalities gotten from \eqref{tsirelsonii} by moving the 
minus sign in front of the other correlators in turn are proved 
similarly and we refer to them all collectively as TsirelsonII.

\subsection{Relationship to Ordinary Quantum Mechanics}

Tsirelson proved  
that TsirelsonII 
are necessary and sufficient conditions for the 
existence of what we will call an ``ordinary quantum model
for the correlators'' (OQMC) \cite{Tsirelson:1985, Tsirelson:1993}. 
An OQMC consists of a vector 
$|\psi>$ in a Hilbert space, and two pairs of self-adjoint operators
$\{S^a, S^{a'}\}$ and $\{S^b, S^{b'}\}$ such that $S^\alpha$ 
commutes with $S^\beta$, for all $\alpha$ and $\beta$, 
and such that 
\begin{equation*}
X_{\alpha\beta} = <\psi| S^\alpha S^\beta |\psi> \;, 
\end{equation*}
for all $\alpha$ and $\beta$. 

TsirelsonII are necessary but not sufficient for the 
existence of an ``ordinary quantum model
 for the probabilities'' (OQMP) however. Such an OQMP would consist of a 
vector $|\psi>$ and two pairs of projective decompositions of unity,
$\{P^+_a, P^-_a\}$, $\{P^+_{a'}, P^-_{a'}\}$,  
$\{P^+_b, P^-_b\}$ and $\{P^+_{b'}, P^-_{b'}\}$, such that
$[P^i_\alpha, P^j_\beta] = 0$ and 
\begin{equation*}
\mathbb{P}_{\alpha \beta}(i j)  = <\psi|P^i_{\alpha} P^j_{\beta}|\psi> \;,
\end{equation*}
for all $\alpha$ and $\beta$, and all $i$ and $j$.
The necessary and sufficient conditions for an OQMP to exist 
are given by Tsirelson in reference \cite{Tsirelson:1980} 
and we refer to them as TsirelsonIII.  

We know that if an OQMP exists then there exists a SPJDF 
\cite{Craig:2006}. It is not known whether the converse is true.  

\section{PR Boxes}

The device that has come to be known as a PR box is 
a box with two knobs -- ``A" and ``B" -- each with two settings
-- $a$ or $a'$ for $A$ and $b$ or $b'$ for $B$ -- which 
may be thought of as inputs. 
For each setting $(\alpha, \beta)$ for the box, it produces two 
outputs, $\pm1$ for $A$ and $\pm1$ for $B$. So the 
setup is simply that of the EPR-Bohm experiment and we 
can employ our notation from above.  The interesting 
features of the PR box are that the probability distributions on the 
outcomes given the settings (i.e. the experimental probabilities
in our previous language) maximally violate one of the CHSHB inequalities
and moreover violate TsirelsonII and so do not admit an OQMC/OQMP.
Noting that each correlator, $X_{\alpha\beta}$ lies between $-1$
and $1$, the ``CHSH'' function $Q \equiv  
X_{ab} + X_{ab'} + X_{a'b'} - X_{a'b'}$ has a ``logical'' upper  
bound of 4.

There is a unique set of experimental probabilities which 
gives rise to correlators 
which saturate this logical bound: $X_{ab} = X_{ab'} = X_{a'b'} =1$
and $X_{a'b'} = -1$. These are 
\begin{align}\label{PRbox.eq} 
\mathbb{P}_{ab}(+1,+1) &= \mathbb{P}_{ab}(-1,-1) = \frac{1}{2}\notag \\
\mathbb{P}_{ab'}(+1,+1) &= \mathbb{P}_{ab'}(-1,-1) = \frac{1}{2}\notag \\
\mathbb{P}_{a'b}(+1,+1) &= \mathbb{P}_{a'b}(-1,-1) = \frac{1}{2}\notag \\
\mathbb{P}_{a'b'}(+1,-1) &= \mathbb{P}_{a'b'}(-1,+1) = \frac{1}{2}\; .
\end{align}
There are 7 other PR boxes, obtainable from this one by 
local operations: Alya can flip each of her inputs or outputs 
and so can Bai \cite{Barrett:2005}. Each PR box maximally 
violates one of
the 8 CHSH inequalities. 

In \cite{Craig:2006} it was shown that the PR box admits a joint 
quantal  measure, indeed a decoherence functional that does the 
job for one of the PR boxes is explicitly given in that paper. 
By doing the appropriate bit-flip on that 
decoherence functional (either flip the outcome 
of the $a'$ measurement or swap $b$ and $b'$: both 
produce the same result), we obtain 
the decoherence functional for the PR box \eqref{PRbox.eq}:
\begin{align*}
{D}_{PR}(-+--;-+--) & = {D}_{PR}(+++-;+++-) = {D}_{PR}(++-+;++-+) \\
& = {D}_{PR}(---+;---+) = \frac{1}{2} \\
-{D}_{PR}(-+-+;-+--) & = -{D}_{PR}(+++-;++--) = {D}_{PR}(---+;----) \\
& = {D}_{PR}(-+-+;+---) = -{D}_{PR}(++-+;+---) \\
& = -{D}_{PR}(---+;+---) = {D}_{PR}(++-+;+++-) \\
& = -{D}_{PR}(++-+;-+-+) = -{D}_{PR}(---+;-+-+) \\
& = {D}_{PR}(---+;++-+) = \frac{1}{4}
\end{align*}
The remaining elements which are not equal to one of the above by
Hermiticity are zero.

Since the correlators
violate TsirelsonII the result of \cite{Craig:2006} 
shows that the measure is not strongly positive but 
it is nevertheless quantal, according to the Sorkin classification. 
The  joint quantal measures for the other PR boxes
are obtained similarly by 
performing 
appropriate bit-flips.

\begin{theorem}
Any system of (no-signalling) experimental probabilities for the (2,2,2) 
setup admits a joint quantal measure.
\end{theorem}
\begin{proof}
A system of experimental probabilities 
for the (2,2,2) setup is a 
vector ${\bf {y}}$ in an eight dimensional real vector space
because there are only eight independent 
probabilities, the rest being fixed by the no-signalling 
conditions and normalisation. ${\bf y}$ 
is an element of a polytope whose 24 vertices, 
${\bf v}_i$, $i = 1, \dots 24$, 
consist of the 8 PR boxes and $2^4 = 16$
so-called ``local'' vertices which 
are the deterministic boxes in 
which each of Alya's and Bai's inputs has
a definite output  
\cite{Barrett:2005}. In other words, ${\bf{y}}$ is a 
convex combination of these vertices,
${\bf{y}} = \sum_{i = 1}^{24} p_i {\bf v}_i$ with 
$\sum_{i = 1}^{24} p_i = 1$ and all $p_i \ge 0$.   

Each of the 8 PR boxes admits a 
joint quantal  measure.

Each local vertex admits a joint  probability (level 1) measure: 
the probability of exactly one of the histories in ${\Omega}$
-- the one that corresponds to the deterministic
box -- is 1 and the other histories have zero  
probability.  A probability measure is
a special case of a joint quantal measure: 
each level of the hierarchy includes the lower levels.

So each vertex of the polytope admits a joint quantal measure 
$D_i(\cdot;\cdot)$
and therefore the measure
$D(\cdot;\cdot) = \sum_{i = 1}^{24} p_i D_i(\cdot;\cdot)$
is a joint quantal measure for
${\bf{y}}$.  
The only condition that might be violated is positivity 
but this
follows from the convexity: a convex combination of 
non-negative numbers is non-negative. 
 
\end{proof}


\section{Discussion}

The question of whether the existence of a SPJQM
implies the existence of a OQMP remains open. One 
could try to find a counterexample by 
finding a set of probabilities on the 
boundary of the region in the polytope 
of probabilities which allows a OQMP. That 
point will admit a SPJQM. One could try to 
find perturbations that take it outside the OQMP
region whilst preserving 
the strong positivity of the decoherence functional.   
The difficulty of this strategy lies in the sheer complexity of 
Tsirelson's conditions for the existence of an OQMP 
\cite{Tsirelson:1980}. 
A counterexample -- a set of experimental probabilities that
does not admit an OQMP but which does admit a SPJDF -- would be 
extremely interesting because it would mean that 
experiment could in principle distinguish between ``ordinary 
quantum theory'' and 
``quantum measure theory plus strong positivity''. 

The question posed by Popescu and Rohrlich and others, 
``Why does nature [apparently] not make use
of the full power of possible non-local non-signalling correlations by 
realising a physical PR box?''\footnote{Ref.\ \cite{Marcovitch:2006}
 claims to
  realize a PR box within ordinary quantum theory by restricting to a final
  state. 
  However, the measurement outcomes no
  longer decohere in this case, and thus the physical 
interpretation of the setup is obscured.}
is transformed, in the framework of generalised measure theory,
into two questions. 
Why does nature (apparently) not make use of level 3
and higher of the hierarchy
of measure theories and why, if it does indeed restrict itself to 
the quantal level, is strong positivity (apparently) 
realised. 

One answer to the former question could be provided if it were
the case that any set of no-signalling probabilities 
for any number of experimenters, settings (depending on 
experimenter)  and outcomes (depending on experimenter and 
setting) admitted a joint quantal measure. Then one could argue that 
nature doesn't use the level 3 measures because there is no need:
one could always concoct a level 2 measure to simulate any
finite set of experimental data. We are sceptical about this possibility
but it needs to be investigated. 
This can be done by studying
correlated probability distributions that have the  
potential of involving irreducible three-history interference. 
Such three-history interference would 
show itself in situations involving either three settings 
for at least one experimenter, or 
three outcomes for at least one experimenter-and-setting.
We suspect that outcomes in such situations
can be correlated in ways that cannot be accounted for at level
2, but can be accommodated at level 3.
For example, we conjecture that the $PR_n$  box
introduced in \cite{Brunner:2006} does not admit a joint 
level $n-1$ measure but does admit a joint level $n$ measure, 
$n \ge 2$. 

Even within the context of level 2 
measures, there remains the question of strong positivity
which is discussed in 
some detail in reference \cite{Martin:2004xi}.
The main reason that the condition is adopted in that paper is 
calculational feasibility. 
Positivity of the 
decoherence functional for the quantal random 
walk is guaranteed by strong positivity which in turn is guaranteed
by positivity of the ``transfer matrix'' which 
propagates the decoherence functional from one time step 
to the next. This is relatively easy to 
satisfy. The condition of positivity on the other hand
is extremely difficult to implement without imposing 
strong positivity.  
At each  
time step there are $2^N -1$ positivity conditions (one for 
each non-empty subset of the sample space)
 to satisfy 
where $N$ is the number of distinct histories to that 
time, and these conditions do not seem to 
translate into any simple condition on the 
transfer matrix. 
Given that the cardinality of the 
sample space of possible paths grows exponentially 
with discrete time, the number of positivity conditions 
grows as an exponential of an exponential in time.

Strong positivity allows the construction of a Hilbert 
space from the decoherence functional. Moreover it 
guarantees that formally combining two non-interacting 
subsystems into a single system does not spoil 
positivity \cite{Martin:2004xi}: if the individual subsystems were 
only positive and not strongly positive, the combined system 
need not be positive.
Are these strong enough {\it physical} arguments for strong 
positivity?

\section{Acknowledgements} 

We thank Jonathan Halliwell, Joe Henson, 
Rafael Sorkin, Jamie Vicary, Hans Westman and the members of the Imperial College Relativity lunch for 
useful discussions.  This work was supported in part by the Marie Curie Research and Training Network ENRAGE (MRTN-CT-2004-005616).

\providecommand{\href}[2]{#2}\begingroup\raggedright\endgroup

\end{document}